\documentclass[showpacs,preprintnumbers,amsmath,amssymb]{revtex4}

\ifx\pdfoutput\undefined
\usepackage{graphicx}
\else
\usepackage[pdftex]{graphicx}
\fi

\usepackage{dcolumn}
\usepackage{bm}

\newcommand{\be}{\begin{equation}}
\newcommand{\ee}{\end{equation}}
\newcommand{\bear}{\begin{eqnarray}}
\newcommand{\eear}{\end{eqnarray}}
\newcommand{\ba}{\begin{array}}
\newcommand{\ea}{\end{array}}

\newskip\humongous \humongous=0pt plus 1000pt minus 1000pt

\newif\ifdtup

\def\oldreffmt#1{\rlap{[#1]} \hbox to 2\parindent{}}

\def\figfmt#1{\rlap{Figure {#1}} \hbox to 1in{}}

\def\beq{\begin{equation}}
\def\eeq{\end{equation}}
\def\bea{\begin{eqnarray}}
\def\eea{\end{eqnarray}}

\def\bq{\begin{quote}}
\def\eq{\end{quote}}

\relax

\newdimen\tdim
\tdim=\unitlength
\def\bar{\overline}

\begin{document}
\setcounter{page}{0}

\begin{flushright}
ANL-HEP-PR-07-96\\ 
EFI/07-31\\
NUHEP-TH/07-11\\
\end{flushright}

\title{Top Compositeness at the Tevatron and LHC }

\author{ {Ben Lillie$^{a, b}$, \ Jing Shu$^{a, b}$, \ 
and Tim M.P. Tait$^{a, c}$}\\[0.5cm]
\normalsize{$^{a}$ Argonne National Laboratory, Argonne, IL 60439}\\
\normalsize{$^{b}$ University of Chicago, Chicago, IL 60637}\\ 
\normalsize{$^{c}$ Northwestern University, 
2145 Sheridan Road, Evanston, IL 60208} }

\begin{abstract}
We explore the possibility that the right-handed top quark is composite.  We
examine the consequences that compositeness would have on $t \bar{t}$
production at the Tevatron, and derive a weak constraint on
the scale of compositeness of order a few hundred GeV from the $t \bar{t}$
inclusive cross section.  More detailed studies of differential properties
of $t \bar{t}$ production could potentially improve this limit.
We find that a composite top can result in an
enhancement of the $t \bar{t} t \bar{t}$ production rate at the LHC 
(of as much as $10^3$ compared to the Standatd Model four top rate).  We
explore observables which allow us to extract the four top rate from the
backgrounds, and show that the LHC can either discover or constrain top
compositeness for wide ranges of parameter space.
\end{abstract}

\maketitle
\thispagestyle{empty} 

\section{Introduction}
\label{sec:intro}

The idea that some or all of the Standard Model (SM) of particle physics may
be composite is a fascinating one, and the discovery of compositeness would
radically redefine most of the fundamental questions for particle physics
today, by mapping the apparent low energy degrees of freedom we see at
the weak scale to a different set in the ultra-violet.  Given the
exciting ramifications of compositeness, it is natural to ask whether or not
we can see signs of compositeness at the LHC, how it could manifest itself,
and what characteristics would distinguish it from other forms of physics
beyond the Standard Model.

The standard search for compositeness looks for higher dimensional
(non-renormalizable) operators \cite{Eichten:1983hw}.  The search for such
operators is a powerful, model-independent means to search for compositeness,
because the low energy effective field theory is not very sensitive to the details
of the theory of compositeness (which at any rate are difficult to estimate
precisely because of the strong couplings involved).  Depending on the
specific operator under consideration, the LHC can discover their presence
up to high scales, of order tens of TeV \cite{:1999fr}.  
However, the effective operator approach also
has its drawbacks.  Effective operators are induced by {\em any} high mass
physics beyond the SM, including weakly interacting possibilities.  To truly
see a theory of compositeness, and recognize it unequivocaly as such, it would
be preferable to see phenomena that can be more specifically associated with
compositeness.

Given the success of the SM in describing experimental data, it's not obvious
that the LHC has much opportunity to see anything beyond the contact
interactions.  Precision measurements from LEP I and SLD put limits between
a few to a few tens of TeV on compositeness of the electroweak gauge or Higgs
bosons \cite{Collaboration:2007ri}.  
LEP II bounds lepton compositeness at tens of TeV \cite{Alcaraz:2006mx}, and the Tevatron
bounds light quark composite operators at between a few and ten TeV 
\cite{Abulencia:2007ez}.  Even the
measurements of left-handed bottom couplings at the $Z$-pole bound 
compositeness of the
third family quark doublet on the order of a few TeV \cite{Alcaraz:2006mx}.  The only sector of the
SM which is currently not strongly bounded by existing measurements is the
right-handed top quark \cite{Georgi:1994ha}.  We thus choose to explore
compositeness of $t_R$ (and no other sector of the SM) as the most likely
place that a low scale of compositeness might be manifest at the LHC.

Models of compositeness are theoretically challenging, because the strong coupling renders 
them difficult to analyze.  Nonetheless, we can proceed using effective field theories, with our
ignorance of the underlying strong dynamics parameterized in terms of coefficients of 
operators whose size we can estimate up to order one uncertainty in terms of
naive dimensional analysis (NDA)  \cite{Manohar:1983md}.  While there are  specific models in
which the top is composite, such as the dual conformal field theory
(CFT) interpretation \cite{Arkani-Hamed:2000ds} of Randall-Sundrum (RS)
\cite{Randall:1999vf} models with gauge fields in the bulk \cite{Davoudiasl:1999tf},
and supersymmetric constructions \cite{Strassler:1995ia}, 
these models invariably postulate that the Higgs and/or left-handed third family quark doublet
are also composite, and thus have compositeness scales probably too high to be probed by the LHC
(however, see \cite{Suzuki:1991kh}).
We choose to work generically in a framework in which {\em only} $t_R$ is composite, without
getting attached to any specific model.  Our hope is to identify interesting phenomena and features
which are not specifically linked to any particular model, but might reasonably be expected to occur
in a broad class of models in which the top is composite.

We begin in Section~\ref{sec:model} by introducing the operators describing
the lowest energy consequences of $t_R$ compositeness.  We place bounds on the scale of
top compositeness by considering the effects of such
operators on the $t \bar{t}$ production rate at the Tevatron in Section~\ref{sec:ttbar},
also finding observables which may improve the analysis in the future.  In
Section~\ref{sec:fourtop} 
we go beyond the operator level, and consider some of the
higher resonances which might accompany a composite $t_R$.  In
Section~\ref{sec:conclusions} we conclude with some outlook.

\section{Top Compositeness: A Bottom-up View}
\label{sec:model}

The first question that arises when one contemplates a composite top is: what is it made of?
We imagine that there is some new force which confines at an energy scale hopefully
accessible to the LHC.  Above the scale of confinement, there should be a weakly coupled
description in terms of a set of constituents (preons), with the SM gauge interactions forming
part of the unbroken non-anomalous chiral symmetries of the new strong force.  Below the
scale of confinement, the physics is described by an effective 
field theory containing the bound states that result, with the right-handed top 
among the lightest of the bound states of this new sector.   Generally, one expects that
confining theories break chiral symmetries and result in massive composite fermions
\cite{Vafa:1983tf}, however one can engineer massless fermions by 
combining 't Hooft anomaly-matching \cite{thooft}
with some inspired model-building \cite{Georgi:1985hf}.
There may
be additional light states (which may or may not themselves be particles
familiar from the Standard Model), and their existence would help pin down the
underlying chiral symmetries of the new confining force.  To minimize
model-dependence we concentrate our focus on the consequences for
observables involving top quarks.

Using the language of effective field theory, we can parameterize the
residual effects of the strong dynamics on the top quarks at the lowest energies.  
The residual effects represent the deviations from point-like behavior of top, and can be
represented at the lowest energies as higher dimensional operators, whose coefficients
we estimate up to order one uncertainties using NDA \cite{Manohar:1983md}.
The largest of these operators is a four-point interaction of $t_R$.
Up to color structures, there is a unique Lorentz-invariant operator
at dimension six which involves only the right-handed top 
quark,
\bea
\frac{g^2}{\Lambda^2} \left[ \bar{t}^i \gamma^\mu P_R t_j \right] 
\left[ \bar{t}^k \gamma_\mu P_R t_l \right]
\label{eq:4top}
\eea
where $\gamma^\mu$ are the Dirac gamma matrices, $P_R$ is the right-chiral
projector, and $g^2 / \Lambda^2$ is the coupling of this new interaction.
It can be understood that $g / \Lambda$ represents the amplitude to create the
composite field, and $\Lambda$ itself characterizes the energy scale at which
further elements of the composite sector become important.  The effective theory
is sensible provided $g \lesssim 4 \pi$.
There are several
possibilities to construct $SU(3)_C$ gauge-invariant combinations of the
color indices $i$, $j$, $k$, and $l$.  Since the Lorentz structure is
suggestive of (the low energy limit of) a massive vector exchange, we
consider only color structures which pair $i$ with $j$ and $k$ with $l$.  The
two options are contractions of two octets $( T^a )^{j}_i ( T^a )^{l}_k$
or two singlets $\delta^j_i \delta^l_k$.
Note that operators involving $c_R$ and $u_R$ are also
possible, and could lead to more stringent bounds from flavor-violating
processes.  
By ignoring such operators we are explicitly making assumptions
about the flavor structure of the UV theory.

At scales of order the confinement scale of the new force, we might expect 
to see resonances which couple strongly to $t_R$.
The precise spectrum of these resonances is more model-dependent, but
we can infer from the fact that $t_R$ was produced as a low-lying bound
state that the preons carry both hypercharge and color, and thus we can 
generically expect that the resonances do as well, which is significant
for the LHC because it implies large production cross sections for
these new states.  Of course, it may be that the strong dynamics is 
not described by any moderately coupled resonances, and the transition to the
fundamental degrees of freedom is quick and complicated.

If there are resonances we can describe with an effective theory, we
can hope that the low energy residual of the strong interaction parameterized
in Eq.~(\ref{eq:4top}) is dominantly produced by among the lightest of
these higher resonant
states, in analogy with vector meson dominance familiar from QCD.  
In that case, we can expect some massive vector particles
(which we will refer to as $\rho_{\mu}$, 
in analogy with QCD) transforming either as an octet or
a singlet under $SU(3)_C$.  From Eq.~(\ref{eq:4top}), we can identify
(up to ${\cal O}(1)$ coefficients) 
$\Lambda$ as the mass of the $\rho$, and $g$ its (large) coupling to right-handed
top quarks.  
Generally,
we can expect a small coupling to light quarks will be induced, 
but its size is model-dependent and
for simplicity we ignore this possibility.
The KK gluon in an RS model (in the dual interpretation) is an example of such a state in a theory
which the induced coupling to light quarks is non-negligible \cite{Agashe:2006hk}.

Far above the confinement scale of the new force, the physics
depends very sensitively on the details of the new interaction and the preons
which experience it.  
Provided the confinement scale is low enough, one can imagine that it might
be possible for the LHC to explore the region where the new force is weak
enough that a perturbative description in terms of the preons themselves
would be appropriate.
The chance to see the preons directly would be the most clear signal of compositeness one
could hope for, and would reveal a lot about the underlying strong dynamics which produced
the top as a bound state.
Carrying the analogy with QCD further, it is  easy to imagine interesting and exotic
phenomena in analogy with ordinary 
QCD, such as production of preons followed by
``showering" under their new strong force, leading to their eventual 
``hadronization'' into a jet of top quarks.  We leave such exploration for future
work.

\section{Top Pairs at the Tevatron}
\label{sec:ttbar}

Before exploring how one might discover top compositeness at the LHC, it
is worthwhile to consider the bounds on top compositeness coming from 
LEP/SLD and the Tevatron run II.
The natural scale of contributions to the precision observables at LEP
was estimated in \cite{Georgi:1994ha}, and the resulting 
bound is typically weaker than is expected from $t \bar{t}$
when only the right-handed top participates in the new strong interactions.

\begin{figure}
\includegraphics[angle=0,scale=0.4]{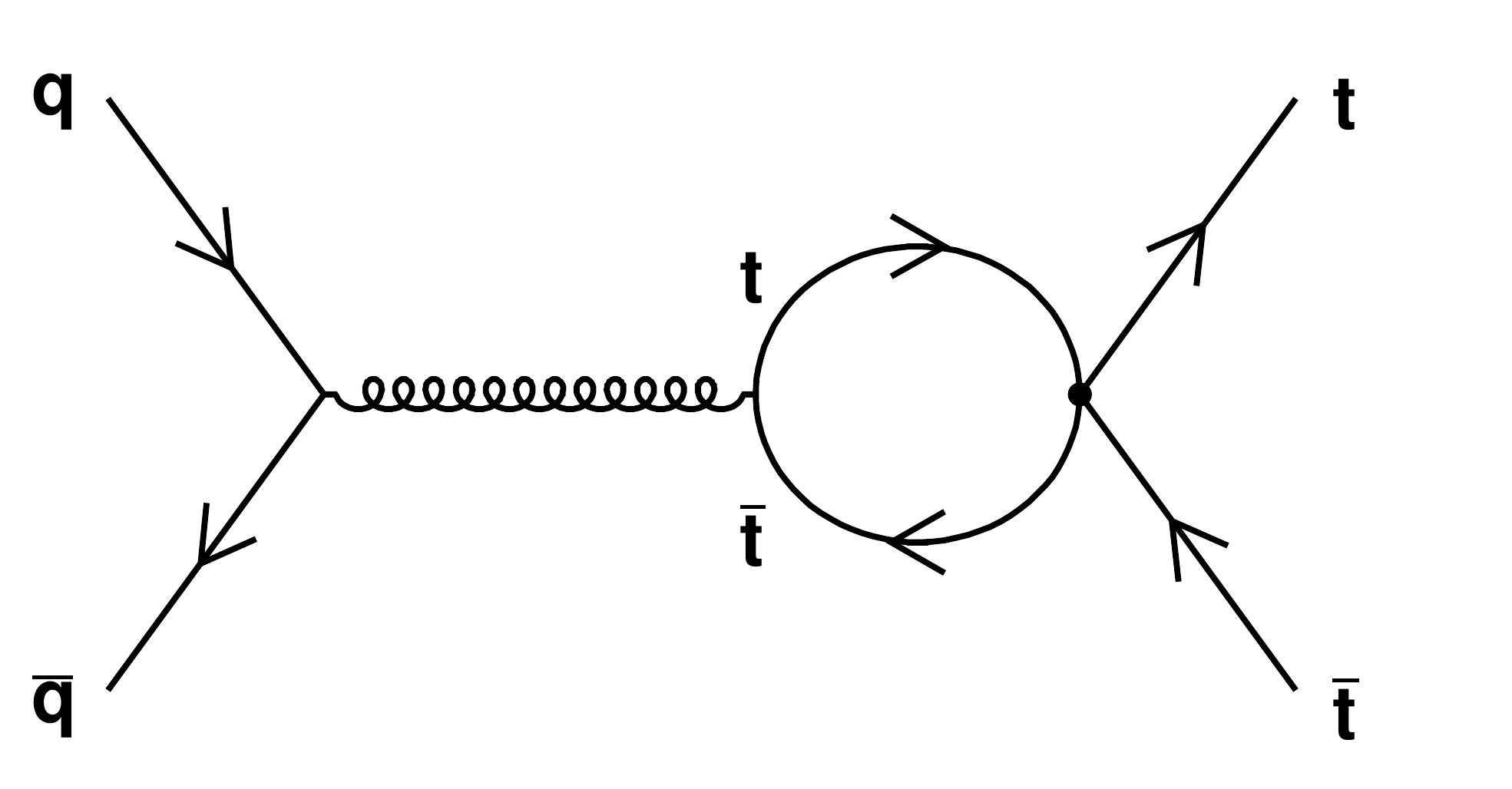} 
\caption{\label{fig:ttbar}
Representative Feynman Diagram showing how the four top quark operator can
contribute to $q \bar{q} \rightarrow t \bar{t}$ at one loop.
}
\end{figure}

At the Tevatron, phase space renders the rate for production of four tops
vanishingly small, and thus we consider the operators which describe the
modification of the top's coupling to gluons resulting from its compositeness.
At dimension six, there are two independent operators involving $t_R$ which
contribute at tree level \cite{Buchmuller:1985jz},
\bea
\frac{g_S}{\Lambda^2}  \left\{ g_1
\left[ \left( H \bar{Q}_3 \right) \sigma^{\mu \nu} \lambda^a P_R t \right] G^a_{\mu \nu}
+ g_2 \left[ \bar{t} \gamma^\mu \lambda^a D^\nu P_R t \right] G^a_{\mu \nu}
\right\}
\eea
where NDA provides the estimate the estimates $g_1(\Lambda) \sim 1/g$ and 
$g_2(\Lambda) \sim 1$.
The first operator is a chromo-magnetic moment for the top \cite{Atwood:1994vm}.
The second operator will be induced by the four top operator of Eq~(\ref{eq:4top})
through renormalization.  When $\Lambda$ is large compared to the energies of
interest, the dominant contribution to $g_2$ will be from the $\log$-enhanced term.
We therefore proceed by inserting the operator of Eq~(\ref{eq:4top}) into a one loop
correction to $t \bar{t}$ production as shown in Figure~\ref{fig:ttbar}.

We know that measurements of the
inclusive $t \bar{t}$ cross section \cite{CDF-tt,D0-tt}
are in rough agreement with the SM predictions \cite{Kidonakis:2003vs},
and therefore 
we expect the data will limit the size of the coefficient 
$g^2 / \Lambda^2$, and thus provide a lower limit on the confinement scale
of the strong dynamics responsible for binding
$t_R$.  We consider the leading effect,
in which the graph of Figure~\ref{fig:ttbar} interferes with the 
tree level Standard Model graph for $q \bar{q} \rightarrow t \bar{t}$ through
a virtual gluon.  We neglect the gluon-initiated graph, which at the Tevatron
amounts to an error of roughly $10\%$ or so in our estimates.
The physical picture behind this process is ordinary production of a pair
of top quarks through the usual strong interaction, followed by their
subsequent re-scattering through the new strong force.

Keeping only the $\log$-enhanced piece, we find that its
contribution to the partonic cross section is proportional to the Standard
Model one,
\bea
\hat{\sigma} (q \bar{q} \rightarrow t \bar{t}) &= &
\hat{\sigma}_{SM} (q \bar{q} \rightarrow t \bar{t}) \times 
\left\{ 1 + c \frac{g^2}{(4\pi)^2} \frac{s}{\Lambda^2} 
\log \left( \frac{\Lambda^2}{m_t^2} \right)
\right\}
\label{eq:sigmattbar}
\eea
where $s$ is the usual Mandelstam invariant corresponding to twice the center
of mass energy of the $t \bar{t}$ pair, 
and $c$ is a coefficient which contains the
color factors, and depends on whether the four-top operator is included in its
the color singlet or color octet version,
\bea
c = + \frac{4}{3} & ~~~~ & ({\rm color~singlet}) , \\
c = - \frac{4}{9} & ~~~~ & ({\rm color~octet}) .
\eea
In deriving Eq.~(\ref{eq:sigmattbar}) we have chosen the renormalization
scale to be $\mu_R = m_t$.
Eq.~(\ref{eq:sigmattbar}) implies that the leading modification is in the
distribution of the center of mass energy of the $t \bar{t}$ system.  
Subleading (non-$\log$ enhanced) terms can also modify the other kinematic 
distributions.

\begin{figure}
\includegraphics[angle=0,scale=0.7]{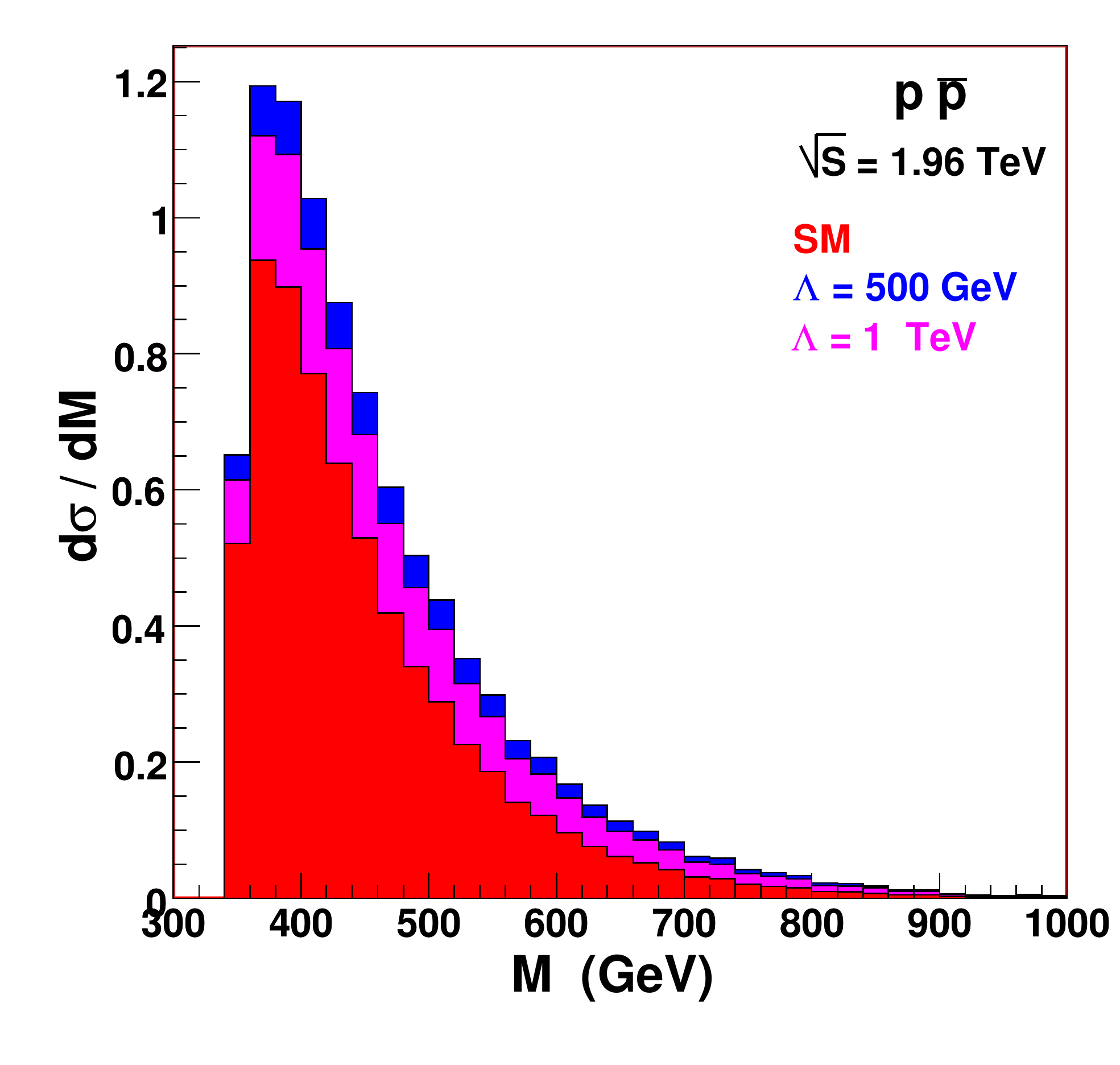} 
\caption{\label{fig:Msinglet}
Invariant mass distribution of $p \bar{p} \to t\bar t$ at the 
Tevatron run II, both in
the Standard Model, and including the singlet-mediated 
contact interaction with $g = 4\pi$, for
$\Lambda = 500$~GeV and $\Lambda = 1$~TeV.
}
\end{figure}

In Figure~\ref{fig:Msinglet} we illustrate the color singlet case,
showing the differential distribution $d\sigma/dM_{t \bar{t}}$ in
terms of the center of mass energy of the top pair system, for
two choices of $\Lambda$ and $g = 4 \pi$.  The behavior is a larger 
increase relative to the SM at higher energies, typical of higher dimension
operators.  The SM cross section (and
convolution with the PDFs) is generated at tree level by the MadEvent code
\cite{Alwall:2007st}.  For up-to-date predictions for the SM $M_{t \bar{t}}$ distribution,
with comparison to different manifestations of new physics in $t \bar{t}$, see
\cite{Frederix:2007gi}.

We expect that the best limit on $\Lambda$ should come from
comparing the $M_{t \bar{t}}$ distribution with data, and we encourage
the experimental collaborations to perform such a fit (which is
very similar to the already-extant search for $t \bar{t}$ resonances
\cite{Aaltonen:2007dz}).  
We are unable to do such a comparison, because the data
with the necessary efficiencies and backgrounds unfolded is not publicly
available.  However, we can compare the effect on the inclusive cross
section to get a rough limit on the size of $\Lambda$.  The inclusive
$t \bar{t}$ cross section is measured by CDF 
(combining several channels) \cite{CDF-tt}
and D0 \cite{D0-tt} to be
\bea
\sigma(t \bar{t})_{CDF} = 7.3 \pm 0.5 \pm 0.6 \pm 0.4~{\rm pb} & ~~~~~~~~ & 
\sigma (t \bar{t})_{D0} = 8.3^{+0.6}_{0.5}~^{+0.9}_{-1.0}~\pm 0.4~{\rm pb}
\eea
(quoted at $m_t = 175$ GeV) where the errors are (in order)
statistical, systematic, and arising from the luminosity measurement.
Both are slightly higher than the Standard Model prediction 
\bea
\sigma (t \bar{t})_{SM} & = & 6.6 \pm 0.8~{\rm pb}
\eea
(we combine results from both references of \cite{Kidonakis:2003vs},
to obtain this estimate), but not 
significantly so.  The CDF measurement has slightly smaller error bars,
and is slightly closer to the SM,  and thus results in the stricter bound.
In order to be conservative, we base our limit on it, combining
the various errors in quadrature to arrive at
$\sigma_{exp} = 7.3 \pm 0.85$~pb.

Because the data are already slightly higher than the SM theory prediction,
and the error bars both experimentally and on the
theory prediction are moderately large, the resulting bound is very weak.
At one sigma, the data require
\bea
\frac{\Lambda}{g} & \gtrsim & 80~{\rm GeV} .
\label{eq:ttbound}
\eea
This is actually low enough that the $\log$-enhanced piece is not necessarily
enhanced compared to the non-$\log$ terms,
and motivates a more careful analysis.  It also is low enough that
even at the Tevatron, the four top operator may not be sufficient to
describe the physics of top compositeness, with large corrections from the
underlying theory in the UV.  For our purposes, to derive a rough bound
on the potential scale of top compositeness, it is sufficient to allow us
to infer that a scale of top compositeness of order a few hundred GeV is
still allowed by the inclusive $t \bar{t}$ cross section.

If a description in terms of a single resonance is appropriate, 
Eq.~(\ref{eq:ttbound}) provides
a bound on the mass divided by coupling of the new state.  For couplings
which saturate NDA ($g \sim 4\pi$), $M \gtrsim 1$~TeV.

\section{Four Tops at the LHC}
\label{sec:fourtop}

At the LHC, the energy is sufficient to explore top compositeness more
directly.  Clearly, Eq.~(\ref{eq:4top}) will lead to an enhancement of the
rate for $p p \rightarrow t \bar{t} t \bar{t}$ provided there is sufficient
parton luminosity at high enough energies from processes such as
$pp \rightarrow t \bar{t}^*$ followed by 
$\bar{t}^* \rightarrow \bar{t} t \bar{t}$ through an insertion of
Eq.~(\ref{eq:4top}).
In fact, the LHC can explore
energies sufficiently above the lower limit of compositeness that one could
hope to directly observe effects beyond the operator level.  Provided there
are sufficiently narrow resonances with masses $\sim \Lambda$, we can
search for them at the LHC.

\begin{figure}
\includegraphics[angle=0,scale=0.7]{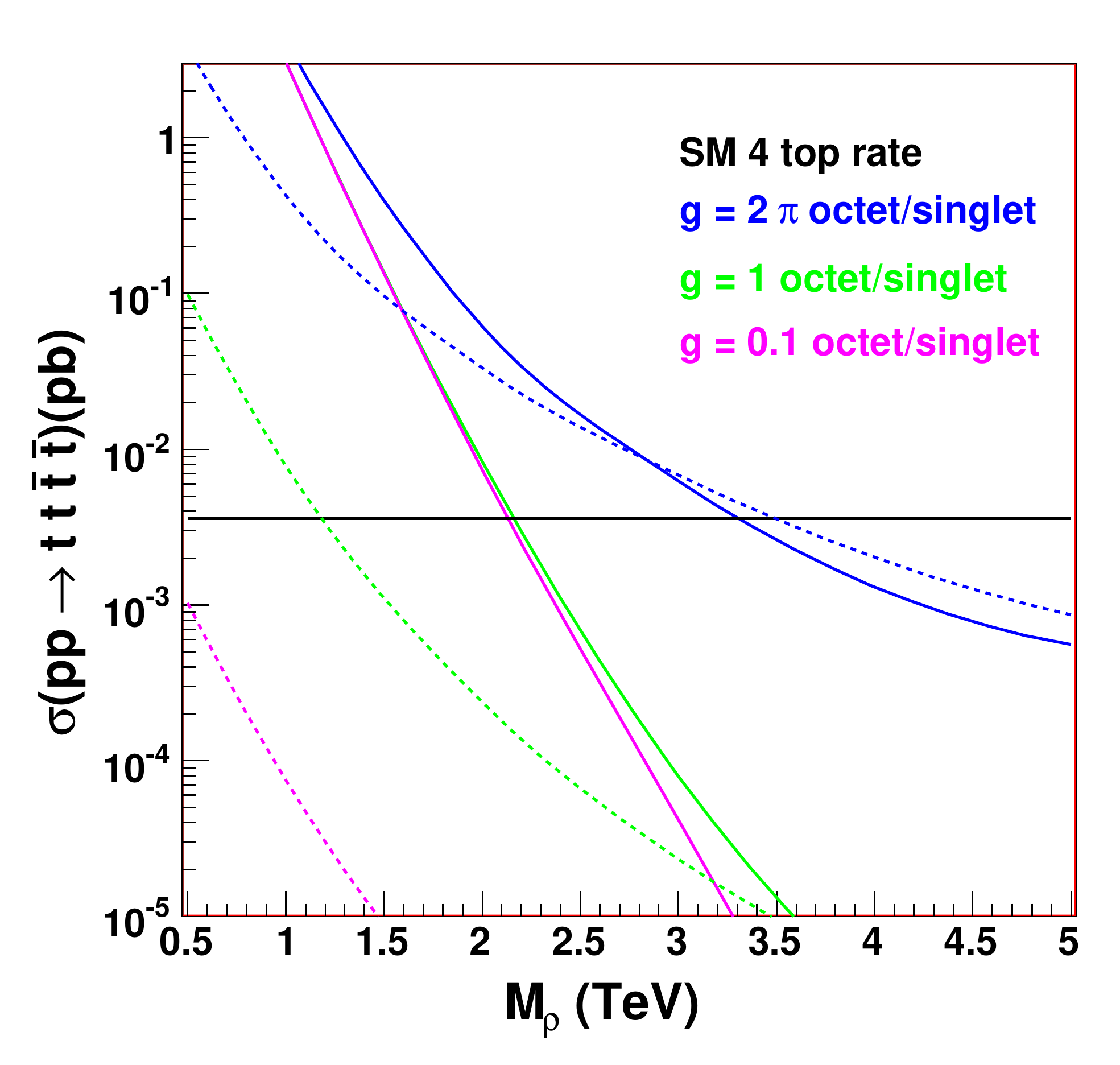} 
\caption{\label{fig:4txsec}
The rate for $t \bar{t} t \bar{t}$ at the LHC as a function of mass $M$
for several values of the coupling $g = 2\pi, 1, 0.1$ (from top to bottom),
for both the case where the
$\rho$ is a color octet (solid lines) or a singlet (dashed lines).
Also shown for reference is the SM 4 top production rate.}
\end{figure}

Thus, we construct an effective theory consisting of the Standard Model plus
a heavy (mass $M$)
vector boson (either octet or singlet), coupled to $t_R$ with
strength $g$,
\bea
-\frac{1}{4} \left( D_\mu \rho_\nu - D_\nu \rho_\mu \right)^2
+ \frac{1}{2} M^2 \rho^\mu \rho_\mu 
+ g \rho_\mu \bar{t}\gamma^\mu P_R t
\eea
where $D_\mu$ is a covariant derivative, containing coupling to gluons
for the octet $\rho$ or not for the singlet $\rho$.
For simplicity, we neglect any coupling to light quarks (in the case
where there are substantial couplings to light quarks, the resonance can
be produced singly through $q \bar{q}$ fusion and the physics is similar
to the KK gauge bosons of a RS model \cite{Agashe:2006hk} ).
At low energies, these new states simply reproduce
the operator of Eq.~(\ref{eq:4top}), whereas at high energies, they can be
resolved as broad (assuming $g \gg 1$) resonances.
We generate events for the reaction $pp \rightarrow t \bar{t} t \bar{t}$ using 
MadEvent \cite{Alwall:2007st}, including
parton showering and hadronization
from PYTHIA \cite{Sjostrand:2006za}, and simulate the detector
using PGS \cite{PGS} with the default LHC detector model.

The inclusive signal rates at the LHC
as a function of $M$ and for several values of the couplings for both color
singlet and color octet $\rho$'s are
shown in Figure~\ref{fig:4txsec}.  Also shown is the SM rate for production
of four tops, which is small by comparison provided $M \lesssim$ a few TeV.  The cases of color octets
and color singlets show a very different dependence on the coupling $g$;
for small $g$ the color octet rate approaches a common value for small fixed
$M$, because the production becomes dominated by the model-independent
rate of $gg \rightarrow \rho \rho$
\cite{Dobrescu:2007yp}, and under our assumptions the branching
ratio for $\rho \rightarrow t \bar{t}$ is one.  For large $M$, it becomes kinematically
favorable to produce a single $\rho$ through $pp \rightarrow t  \bar{t} \rho$, and the dependence on
$g$ is stronger.
The color singlet rate, instead,
is always proportional to $g^2$, because that case always proceeds via
$pp \rightarrow t  \bar{t} \rho$.

Reconstructing all four top quarks is very difficult, suffering
from huge conbinatoric problems. We thus adopt
the simpler signature of at least two like-sign leptons, $\ell^\pm \ell^{\prime \pm}$
plus two hard jets (with $p_T > 20$~GeV and $|y| < 2.5$).
Two well-reconstructed leptons with $p_T > 30$~GeV, $|y| < 2.5$,
are sufficient to trigger,
and demanding like-signs for the leptons severely reduces the physics
backgrounds to processes such as $WZjj$ and $W^\pm W^\pm jj$.  
There is also a contribution from $W^\pm b \bar{b}$
(including single top), with one of the
bottom quarks decaying semi-leptonically.  We reduce this
background with an isolation
cut  \cite{Sullivan:2006hb} around both leptons, requiring each be seperated from the nearest
jet by at least $\Delta R \geq 0.2$.
We also consider ``fake'' backgrounds including
$Wjjj$ where the additional jet fakes a lepton and $W^+ W^- jj$ where one of
the leptons from the $W$ decays is mis-identified to have the wrong charge.
The dominant contribution to this last signature is 
from $t \bar{t}$ production.  

\begin{table}
\begin{tabular}{lcc}
~Process~~~~~   & ~~~Raw Rate~~~  & ~~~After Cuts~~~ \\ 
\hline \\
$W^+ Z jj$      & 6.65 pb         & 1.12 fb \\
$W^- Z jj$      & 4.11 pb         & 0.41 fb \\
$W^+ W^+ jj$    & 0.29 pb         & 0.83 fb \\
$W^- W^- jj$    & 0.13 pb         & 0.32 fb \\
$W^+ b \bar{b}$ &  196 pb         & 0.57 fb \\
$W^- b \bar{b}$ &  136 pb         & 0.18 fb \\ \\
$W^+ W^- jj (t\bar{t})$    &  390 pb         & 3.16 fb \\
$W^+ jjj$       & 2170 pb         & 0.32 fb \\
$W^- jjj$       & 1520 pb         & 0.29 fb \\
\hline\\
Total &   & 7.20 fb 
\end{tabular}
\caption{The background raw event rates, and rates
after acceptance cuts, requiring like-sign
leptons, isolation, and $H_t \geq 1$~TeV.}
\label{tab:cutrates}
\end{table}

\begin{figure}
\includegraphics[angle=0,scale=0.35]{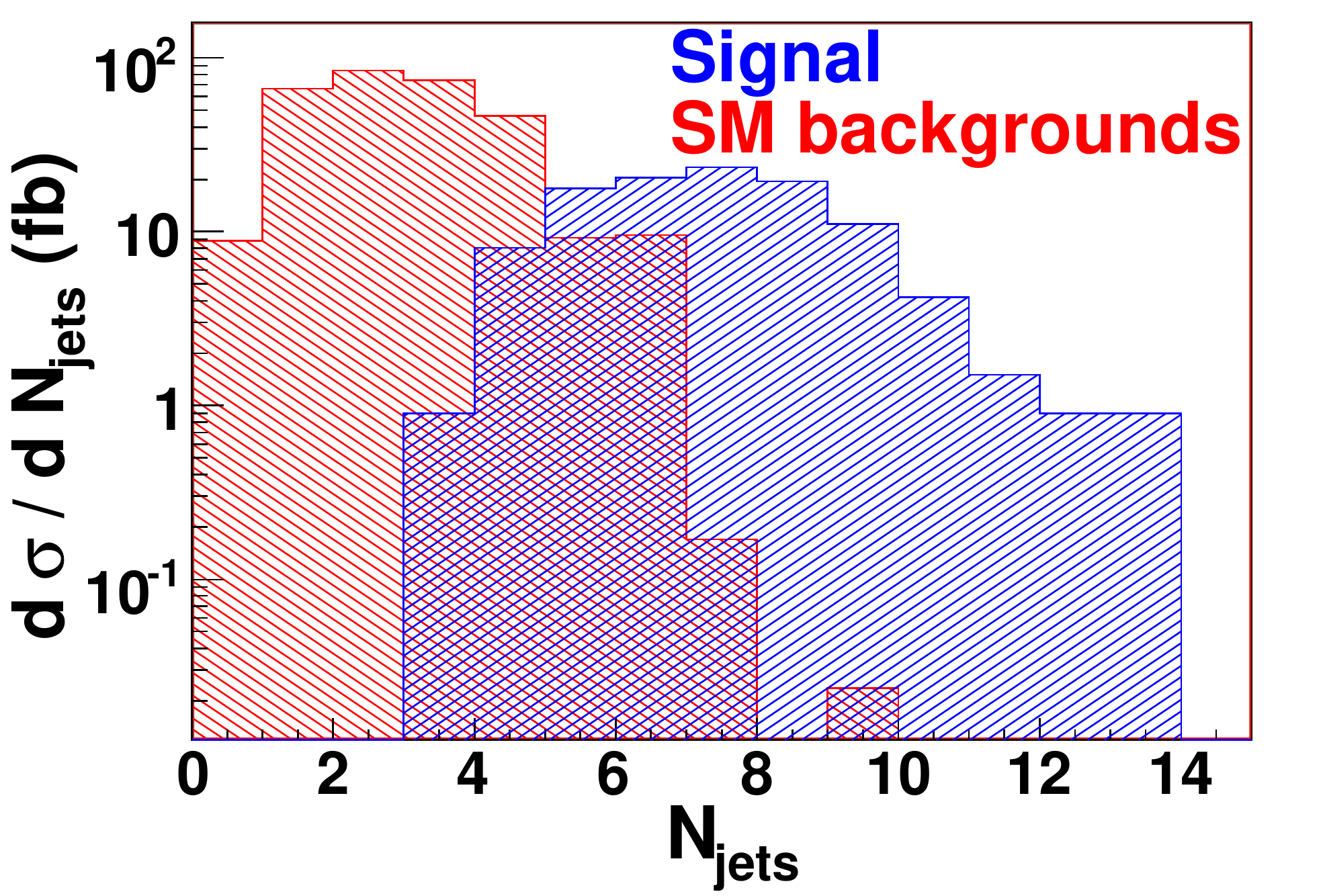} 
\includegraphics[angle=0,scale=0.35]{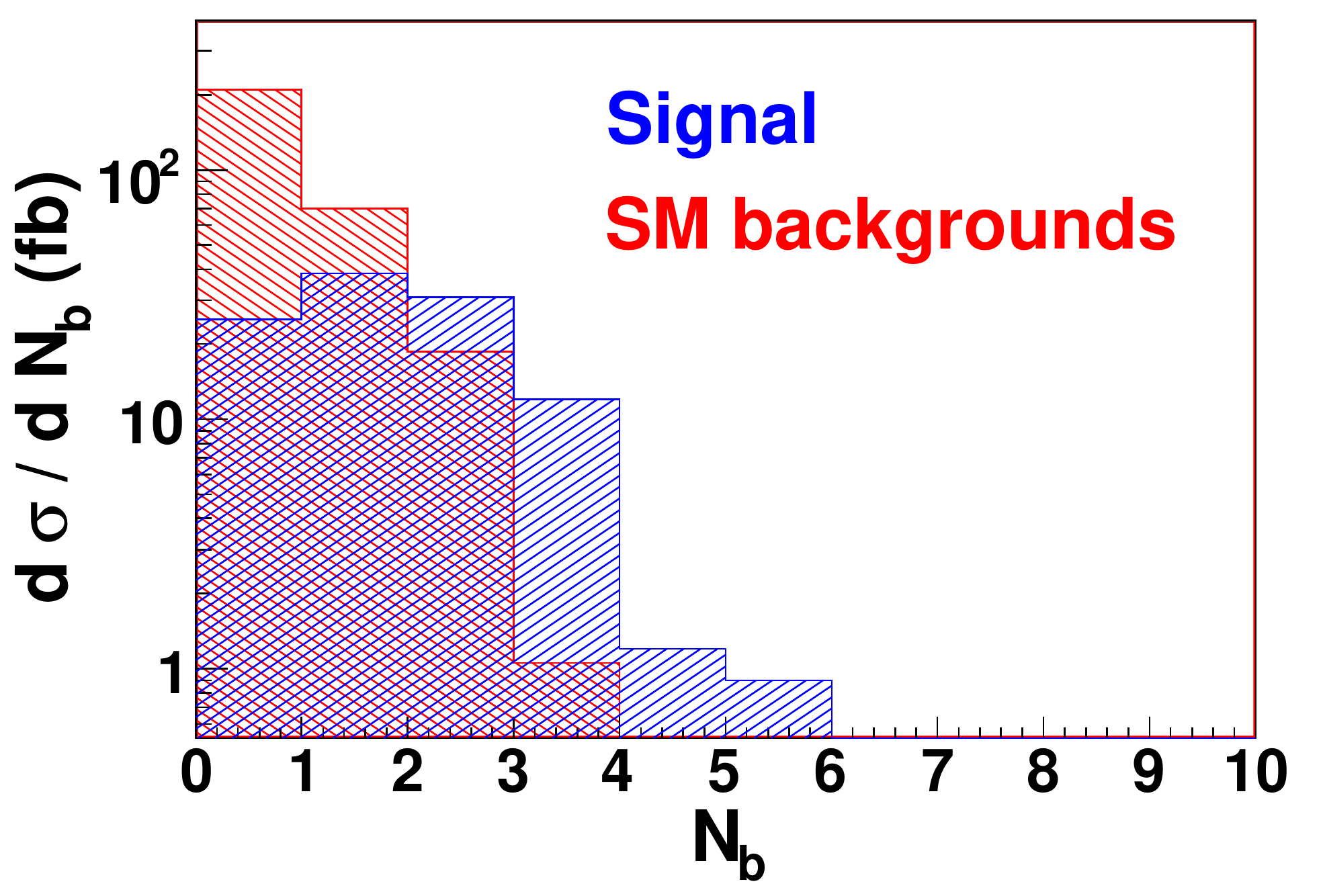} 
\\
\includegraphics[angle=0,scale=0.35]{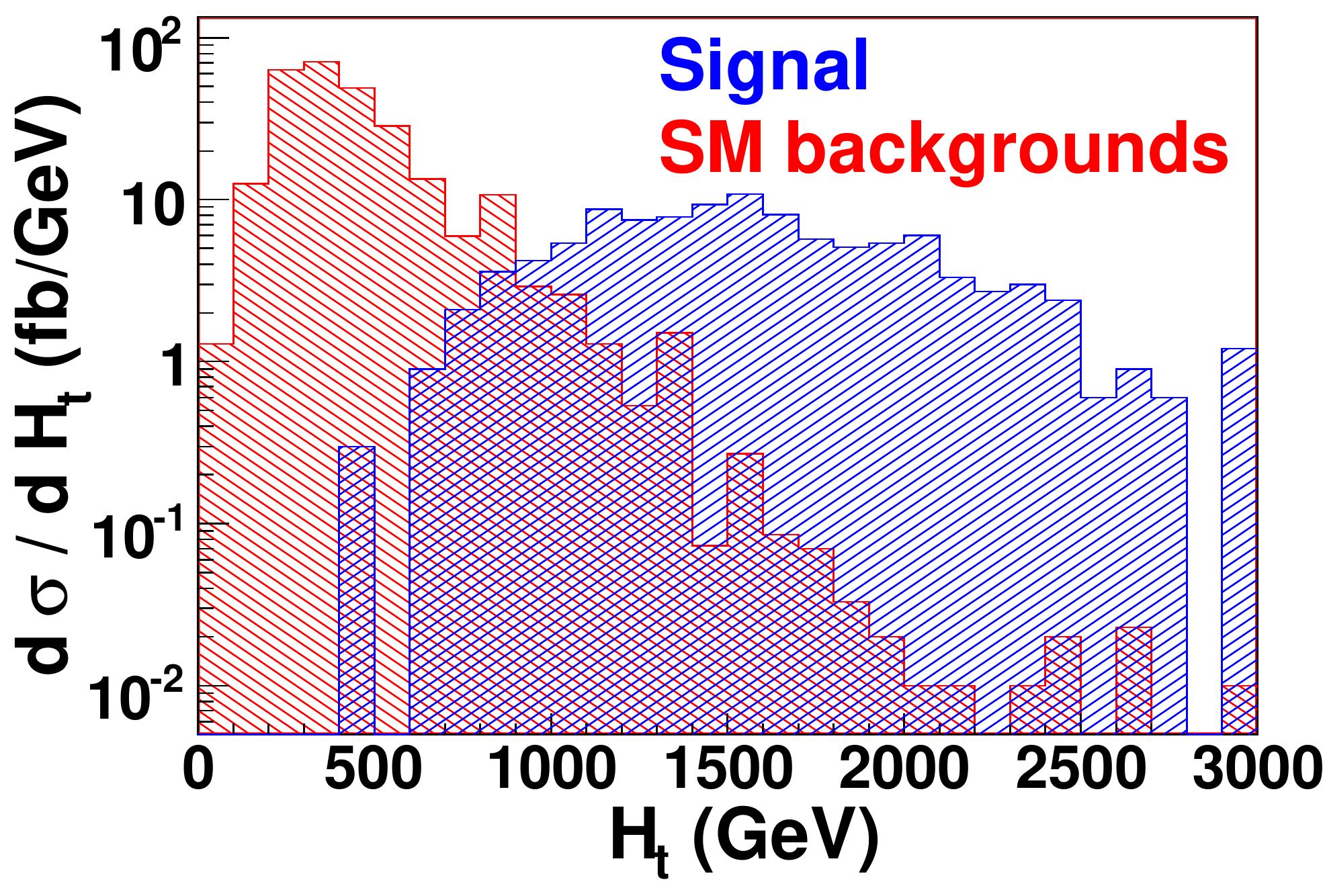} 
\includegraphics[angle=0,scale=0.35]{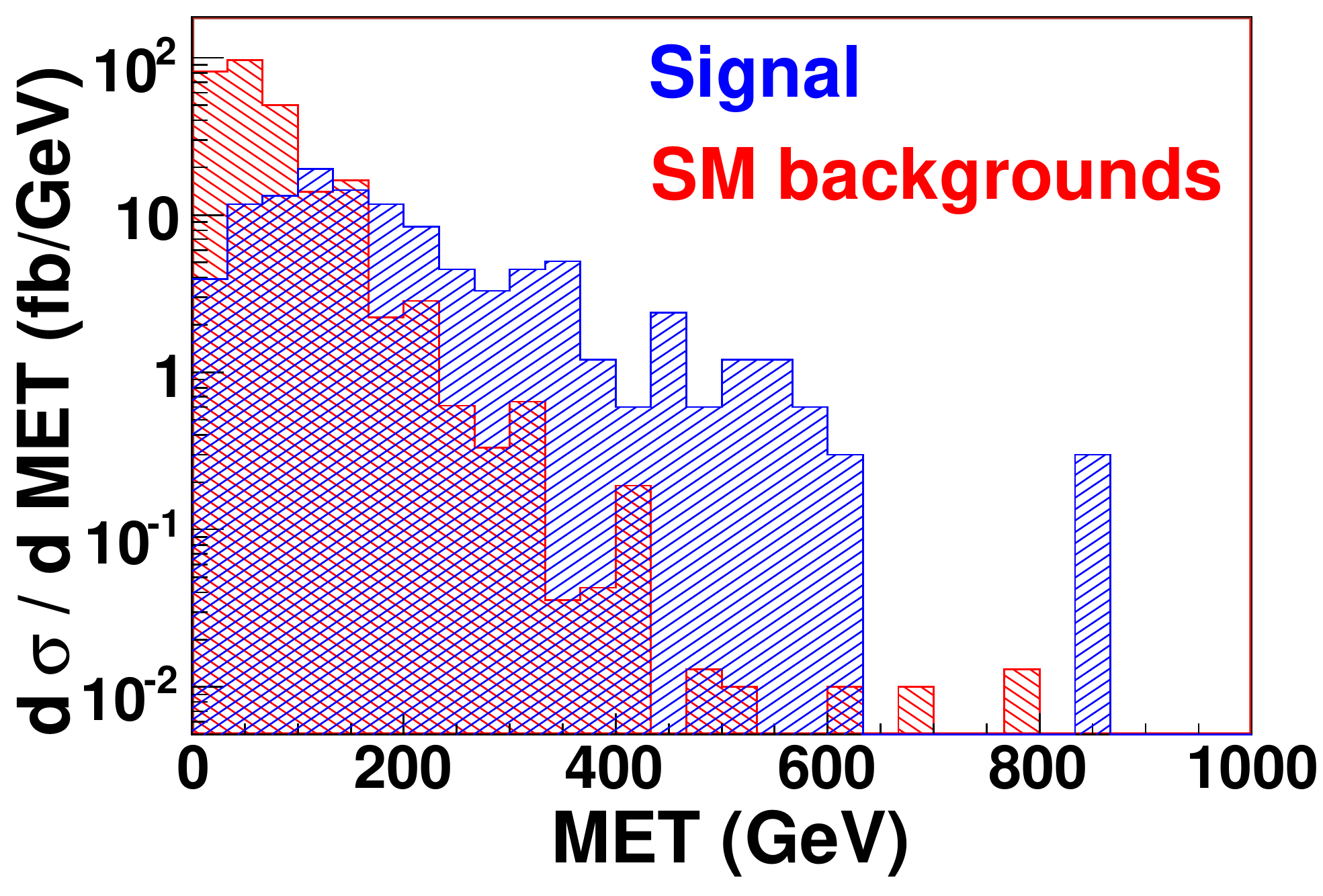} 
\caption{\label{fig:distribution}
Differential cross section for the signal (with $M=1$~TeV and $g=2\pi$) and SM backgrounds
with respect to the number of hard jets (upper left), number of $b$-tagged jets (upper right),
$H_t$ (lower left), and missing transverse momentum (lower right).  The $H_t$ distribution is plotted
before the $H_t$ cut is applied; the other three distributions include all of the cuts.}
\end{figure}

To extract only high center-of-mass energy
events which can correspond to production of four top quarks, we require
$H_t$, defined as the scalar sum of the $p_T$ of all jets, leptons, and
missing transverse momentum satisfies $H_t \geq 1$~TeV.  In 
Figure~\ref{fig:distribution}
we plot the $H_t$ distributions for the signal as well as the sum of the SM
backgrounds.  A cut at 1 TeV dramatically reduces the background (most of which
is from $t \bar{t}$) while only modestly reducing the signal.
We begin with
these simple criteria, and then consider some additional variables
which can dramatically help argue for the ``four top-ness'' of the events
below.  In \cite{Gerbush:2007fe}, it was argued that one could also attempt to
reconstruct the top quarks directly.  One could attempt their procedure
either after our choice of signal analysis, or instead of it, but we
restrict ourselves to the more conservative choice of like-sign leptons and
two hard jets outlined above.
The signal
acceptance is roughly $3\%$, most of which comes from the fact that we have
asked two of the $W$'s with the same charges from the top decays to decay
leptonically.  We expect it depends weakly on the $\rho$ mass $M$.

After applying the acceptance, isolation, and $H_t$
cuts, we find the background
processes yield the rates in Table~\ref{tab:cutrates}.  All of the rates
are estimated based on simulations with MadEvent, followed by showering
and hadronization with PYTHIA, and the detector simulation with PGS
(using the default generic LHC detector model).  The exception is the
$Wjjj$ fake rates, which we estimate by applying a $10^{-4}$ probability
that a jet which passes our acceptance cuts can fake a lepton.  If one managed
to reduce the fake backgrounds sufficiently, the $WZjj$ rates could become
significant.  These are reduced by rejecting events where two of the 
leptons reconstruct an invariant mass close to the $Z$ boson mass (87.2 GeV $\leqslant M_{l^+ l^-} \leqslant$ 95.2 GeV).

The backgrounds sum to about 7.2 fb, about half of which
is the fake rate from a lepton whose charge is mis-reconstructed.  After that,
the leading backgrounds are the $Wjjj$ fake rate, and $W b \bar{b}$.
With $100~{\rm fb}^{-1}$ of collected data,
a $5\sigma$ discovery requires a signal cross section greater than
about 1.4 fb after cuts, or a $t \bar{t} t \bar{t}$ production rate greater
than about 45 fb.  From Figure~\ref{fig:4txsec}, we see that this
corresponds to color octet (and strongly coupled
color singlet) $\rho$'s up to about $2 \sim 3$ TeV.  The Standard Model
rate for $t \bar{t} t \bar{t}$ production, on the other hand, is extremely tiny, about
$3.6$ fb (and so is not visible against the background
using our search strategy). Our analysis is conservative, and could potentially be improved by tightening the cuts, such as requiring more hard jets and/or $b$-tagged jets, or attempting to 
reconstruct the four tops, provided this can be done with sufficient efficiency.

We also consider observables which could be helpful to suggest that
an observed signal has been produced by a four top state.  In 
Figure~\ref{fig:distribution}, we plot the differential cross section with respect
to the total number of reconstructed jets.  A four top final state, even
with two like-sign leptons, produces eight hard quarks (including bottom 
quarks).  QCD radiation can increase this further, whereas at large numbers
the jets initiated by the hard partons begin to fill the entire detector,
and can be merged by the jet-finding process.  (The number of jets from
the hard process is also less in the cases with three or four semi-leptonic
decays).  As can be seen in the figure,
the number of jets reconstructed by PGS (for a signal with $M=1$~TeV and
strong coupling $g=2\pi$) does have a broad maximum around $n_{jet}=8$, 
and is quite different than any of the background processes, for which
the hard process produces many fewer hard partons.

Four top quarks also produce four bottom quarks when they decay.  However,
the probability of tagging all four bottoms is rather small.  In 
Figure~\ref{fig:distribution} we plot the number of bottom quarks tagged by PGS.  The
four top signal peaks at something around two to three bottoms being
reconstructed, whereas the background peaks around zero or one (dominated by $t \bar{t}$).
The presence of several bottom tags together with the pair of charged
leptons is highly suggestive of a multi-top intermediate state.  Further,
the balance between the positive versus negative like-sign leptons in the 
signal sample provides another clue.  A four top final state predicts
equal numbers of positive and negative lepton pairs, whereas production
of multi-$W$s through electroweak processes will show more positive lepton
pairs because of the larger number of valence up quarks compared to down
quarks in the protons.

Other models may lead to an excess over the SM in the channel we consider.  For example,
in supersymmetric models one may pair produce gluinos ($\tilde{g}$) which have a decay
chain such as $\tilde{g} \rightarrow t \tilde{t}^* \rightarrow t \bar{t} \tilde{\chi}^0_1$, leading
to a $t \bar{t} t \bar{t} \tilde{\chi}^0_1 \tilde{\chi}^0_1$ final 
state \cite{Hisano:2002xq}.  One can hope the distribution
of missing transverse momentum will distinguish such a supersymmetric signal from a model
of top compositeness.  Another possibility is pair production of a $b^\prime$ (predicted
in models with an extended custodial symmetry to protect $Z \rightarrow b \bar{b}$ from receiving
large corrections \cite{Agashe:2006at}
or to explain the measurement of $A^{FB}_b$ \cite{Choudhury:2001hs}), which decay into
$b^\prime \rightarrow W^- t$, leading to a $W^+ W^- t \bar{t}$ final 
state \cite{Dennis:2007tv}, which results in events typically
containing less $b$-tagged jets.

\section{Conclusions and Further Thoughts}
\label{sec:conclusions}

The possibility that the top is composite is fascinating, would force us
to rethink our picture of physics in the ultraviolet, and may represent the
unique sector of the SM for which the LHC has an opportunity to see constituents.  We have
studied bounds on top compositeness from $t \bar{t}$ production at the
Tevatron, and find that constraints from the inclusive $t \bar{t}$ cross section are weak, though
stronger bounds could potentially be obtained by studying kinematics instead of the total rate.
Some models may have moderately coupled
vector resonances that describe the physics around the compositeness scale.  
Such models generally lead to a large, observable excess in four top quark production at the LHC.
We perform a conservative analysis that searches for an excess of events containing four top
quarks decaying into at least
two like-sign leptons and at least two hard jets,
and find we can observe a $5\sigma$ excess for new states up to about
2 TeV, or more if the new states are colored and/or strongly coupled to the
top quark.

The next step is to examine models with which one can ask questions about
the phenomena most intimately tied to top compositeness, and to determine
whether or not we can see such constituents at the LHC, and perhaps unravel
the difference between different constructions.  One could imagine seeing direct production
of the top constituents, and maybe even ``showering" or ``hadronization" effects of the new strong force, provided the UV theory is accessible.  Models of top compositeness are
challenging to analyze, but they lead to unique phenomena and inspire us to consider top events
with unusual kinematics we might otherwise overlook.  As the LHC turn-on approaches, 
it behooves us to explore them!

{\bf Acknowledgments }

The authors are pleased to acknowledge conversations with J. Alwall,
D. Amidei, E. Brubaker, D. Choudhury, 
B. Dobrescu,
R. Erbacher, K.C. Kong, T. LeCompte, J. Lykken,
R. Mahbubani,
M. Peskin, J. Rosner, M. Strassler, C.E.M. Wagner, and M. Weber.
Research at Argonne National Laboratory is 
supported in part by the Department of Energy 
under contract DE-AC02-06CH11357.


\end{document}